\documentclass[aps,prd,nofootinbib,showpacs,preprintnumbers,longbibliography,amssymb,11pt]{revtex4-1} 
\usepackage[sort&compress]{natbib}

\usepackage{graphicx}
\usepackage{epsfig}
\usepackage{dcolumn}
\usepackage{bm}
\usepackage{amsmath}

\usepackage{slashed}

\graphicspath{{./Figures/}}

\newcommand{\eqnref}[1]{Eq.~(\ref{eqn:#1})}

\newcommand{\secref}[1]{Sec.~\ref{sec:#1}}

\newcommand{\figref}[1]{Fig.~\ref{fig:#1}}

\newcommand{\tableref}[1]{Table~\ref{table:#1}}

\begin{document}

\preprint{MITP-16-040}

\title{Phenomenology of Enhanced Light Quark Yukawa Couplings and the
  $W^\pm h$ Charge Asymmetry}

\author{Felix Yu}
\email{yu001@uni-mainz.de}
\affiliation{PRISMA Cluster of Excellence \& Mainz Institute for
  Theoretical Physics, Johannes Gutenberg University, 55099 Mainz,
  Germany}


\begin{abstract}
I propose the measurement of the $W^\pm h$ charge asymmetry as a
consistency test for the Standard Model (SM) Higgs, which is sensitive
to enhanced Yukawa couplings of the first and second generation
quarks.  I present a collider analysis for the charge asymmetry in the
same-sign lepton final state, $p p \to W^\pm h \to (\ell^\pm \nu)
(\ell^\pm \nu jj)$, aimed at discovery significance for the SM $W^\pm
h$ production mode in each charge channel with 300~fb$^{-1}$ of 14~TeV
LHC data.  Using this decay mode, I estimate the statistical precision
on the charge asymmetry should reach 0.4\% with 3~ab$^{-1}$
luminosity, enabling a strong consistency test of the SM Higgs
hypothesis.  I also discuss direct and indirect constraints on light
quark Yukawa couplings from direct and indirect probes of the Higgs
width as well as Tevatron and Large Hadron Collider Higgs data.  While
the main effect from enhanced light quark Yukawa couplings is a rapid
increase in the total Higgs width, such effects could be mitigated in
a global fit to Higgs couplings, leaving the $W^\pm h$ charge
asymmetry as a novel signature to test directly the Higgs couplings to
light quarks.
\end{abstract}


\maketitle

\section{Introduction}
\label{sec:Introduction}

After the discovery of the Higgs boson in 2012 by the ATLAS and CMS
experiments~\cite{Aad:2012tfa, Chatrchyan:2012xdj}, the experimental
Higgs effort has transitioned to a full-fledged program of Higgs
characterization and precision measurements of its couplings to
Standard Model (SM) particles.  The direct observation of the Higgs to
vector bosons has been established at high
significance~\cite{Aad:2015gba, Khachatryan:2014jba,
  Khachatryan:2016vau}, while decays to taus and bottom quarks have
yet to reach discovery significance and direct knowledge about the
couplings of the Higgs to first and second generation fermions is
utterly lacking.

The most straightforward information about light generation Yukawas
would come from direct decays of the Higgs.  While these are certainly
viable possibilities for the charged leptons~\cite{Aad:2014xva,
  Khachatryan:2014aep}, the inability to distinguish light
quark-initiated jets from each other renders this avenue a practical
impossibility, with the notable exception of charm tagging.  A few
studies~\cite{Delaunay:2013pja, Perez:2015aoa, Perez:2015lra} have
investigated the prospects for identifying direct decays of Higgs to
charm jets, where bottom- and charm-jet tagging work in tandem to
disentangle enhanced bottom and charm Yukawa couplings.

Aside from direct decays of the Higgs to light quark jets, the other
possibilities for measuring light quark Yukawa couplings come from
charm--Higgs associated production~\cite{Brivio:2015fxa}, which also
requires a careful calibration of charm jet tagging efficiencies and a
precise determination of Higgs and associated jet backgrounds.  The
practical applicability of this technique is not well established,
however, since a systematic treatment of Higgs and non-Higgs
backgrounds is still absent.

An enhanced light quark Yukawa can also lead to significant effects in
rare Higgs decays to quark--anti-quark mesons and vector
bosons~\cite{Isidori:2013cla, Bodwin:2013gca, Kagan:2014ila,
  Koenig:2015pha}.  The impressive control of theoretical uncertainty
in these calculations and the corresponding proof of principle
searches for such rare decays from $Z$ and Higgs
bosons~\cite{Aad:2015sda, Khachatryan:2015lga, Aaboud:2016rug} make it
an interesting channel to pursue.  In these channels, though,
interpreting a deviation from the SM expectation would require
knowledge of the Higgs vertices in the so-called indirect
contributions.  A deviation in the rate for $h \to J/\Psi \gamma$, for
example, could be attributed to a nonstandard effective coupling of
the Higgs to two photons as well as the charm Yukawa coupling.  Hence,
the realistic sensitivity of these rare Higgs decays to nonstandard
light quark Yukawas suffers not only from the small expected SM rates,
but also because the indirectness of the probe necessitates a
combination with other Higgs measurements.

Nevertheless, the power of combined fits to Higgs signal strengths
cannot be discounted as an important tool in constraining nonstandard
Yukawa couplings~\cite{Perez:2015aoa, Kagan:2014ila, Meng:2012uj}.
Such combined fits, however, are handicapped by the inability to
determine the total width of the Higgs and thus require
model-dependent assumptions in order to extract Higgs
couplings~\cite{Dawson:2013bba}.  For example, the possibility of
exotic production modes of the Higgs boson contaminating the Higgs
dataset~\cite{Yu:2014mda} would introduce new physics parameters
outside of the coupling deviation framework, spoiling the entire
applicability of the $\kappa$-framework.

We see that many of the proposed tests of non-standard Yukawa
couplings have varied difficulties in experimental applicability or
theoretical interpretation.  While direct decay tests are best and
subject to the least theoretical bias, the only potentially viable
channel is the $h \to c\bar{c}$ decay.  Production tests, like
measuring $hc + h\bar{c}$ production, are fraught with many
backgrounds and experimental challenges such as charm tagging.
Indirect tests, whether via Higgs rare decays to quantum
chromodynamics (QCD) mesons and vectors or combined coupling fits to
Higgs data, are most robust when conducted as consistency tests of the
SM.

In the spirit of offering new channels for probing the Standard Model
Yukawa couplings, we motivate the charge asymmetry in vector boson
associated Higgs production at the LHC.  As a proton-proton machine,
the LHC handily favors $W^+ h$ production over $W^- h$ production,
mainly through the Higgsstrahlung process $q q' \to W^{\pm*} \to W^\pm
h$.  At the 14~TeV LHC, for example, with $m_H = 125.09$~GeV,
$\sigma(W^+ h) / \sigma(W^- h) = 1.56$~\cite{YellowReport,
  Heinemeyer:2013tqa}.  We point out, however, that this inclusive
charge asymmetry is dramatically changed if the light SM quarks have
large Yukawa couplings.  Concomitant effects from large light quark
Yukawa couplings, such as $q\bar{q}$ $s$-channel Higgs production and
a rapid increase in the total Higgs width, provide additional channels
for indirectly constraining enhanced quark Yukawas.

In~\secref{Yukawas}, we provide a theory motivation and background on
Yukawa coupling deviations.  In~\secref{charge}, we discuss the charge
asymmetry of $pp \to W^\pm h$ production in the SM and the
modifications induced by anomalous light quark Yukawa couplings.  We
then present a collider analysis for same-sign leptons targetting the
$W^\pm h$ charge asymmetry measurement in~\secref{collider},
demonstrating that the charge asymmetry can be measured at the LHC to
subpercent accuracy.  We proceed to discuss other phenomenological
consequences of enhanced light quark Yukawa couplings and their
constraints in~\secref{signalstrength}.  We conclude
in~\secref{conclusions}.

\section{Yukawa deviations}
\label{sec:Yukawas}
The question of fermion mass generation is a central aspect of the
structure of the Standard Model.  A nonstandard Yukawa coupling in the
SM Lagrangian leads to unitarity violation for $f \bar{f} \to VV$
scattering amplitudes.  In the Higgs post-discovery phase, and in the
absence of direct knowledge of the Yukawa coupling for a given SM
fermion $f$, we can calculate a unitarity bound from $f \bar{f} \to
W^+ W^-$ scattering~\cite{Appelquist:1987cf} by requiring the partial
amplitude satisfies unitarity, $|a_0| \leq 1/2$.  The scale of
unitarity violation is then given by
\begin{equation}
E_f \simeq \dfrac{8 \pi v^2 \xi}{|m_f - y_f v|} \ ,
\label{eqn:unitarity}
\end{equation}
where $v = 246$~GeV is the Higgs vev, $\xi = 1/\sqrt{3}$ for quarks
and $\xi = 1$ for charged leptons.  This unitarity violation is a
general feature in theories with chiral fermion masses arising from
spontaneous symmetry breaking if the fermion mass is mismatched with
its Yukawa coupling.  A stronger bound on $E_f$ can be found by
studying $f \bar{f}$ scattering to arbitrary numbers of longitudinal
modes of electroweak bosons~\cite{Dicus:2005ku}.  

Resolving the mass-Yukawa coupling mismatch necessarily requires
either new sources of $SU(2)_L$ breaking beyond the Higgs vacuum
expectation value (vev) or new matter fermions which mix with the SM
fermions.  Such completions would add new diagrams to the partial wave
amplitude calculated above in precisely the necessary manner to remove
the $\sqrt{s}$ growth in the amplitude.

We note that regardless of the source of the new sources of Yukawa
deviations, the unitarity bound can be far beyond the reach of the
LHC.  For example, light quarks with $\mathcal{O}(1)$ Yukawa couplings
(which requires fine-tuning of SM and new physics Lagrangian
parameters to reproduce the physical light quark masses) motivate $E_f
\sim 3.6$~TeV as the scale of unitarity breakdown.  Although such a
fine-tuned light quark mass is aethestically unappealing, such a
mismatch between the quark mass and the Higgs Yukawa coupling cannot
be discounted from collider searches for heavy fermions, seeing that
limits on vector-like top parters reach only the $1$~TeV
scale~\cite{Aad:2015kqa, Khachatryan:2015oba}.

The unitarity bound and inadequacy of the ad-hoc renormalizable
Lagrangian can be simultaneously cast into more familiar language by
appealing to dimension-6 effective operators for Higgs physics.  Here,
the SM provides the usual dimension-4 couplings that preserve the
mass-coupling relation expected in SM physics, but the fermion masses
and their Yukawa couplings get additional contributions from
dimension-6 operators.  We have
\begin{eqnarray}
\mathcal{L} &\supset& y_u \bar{Q}_L \tilde{H} u_R + y_u' \frac{H^\dagger H}{\Lambda^2} \bar{Q} \tilde{H} u_R \nonumber \\
&+&  y_d \bar{Q}_L H d_R + y_d' \frac{H^\dagger H}{\Lambda^2} \bar{Q} H d_R + \text{ h.c.} \ ,
\label{eqn:Lagrangian}
\end{eqnarray}
where $y_u$, $y_u'$, $y_d$ and $y_d'$ are $3\times 3$ matrices in the
flavor space of $Q_L$, $u_R$, and $d_R$.  The flavor rotations of $Q_L
= (u_L \ d_L)^T$, $u_R$, and $d_R$ are then used to ensure the mass
matrices,
\begin{equation}
m_f = \frac{y_f v}{\sqrt{2}} + \frac{y_f' v^3}{2 \sqrt{2} \Lambda^2} \ ,
\end{equation}
are diagonal, with $f$ denoting up-type or down-type quarks, and we
have expanded $H = \frac{1}{\sqrt{2}} (h + v)$ about its vev.
Importantly, these flavor rotations does not guarantee in general that
the Yukawa matrices
\begin{equation}
\dfrac{y_{f, \text{ eff}}}{\sqrt{2}} = \frac{y_f}{\sqrt{2}} + \frac{3y_f' v^2}{2 \sqrt{2} \Lambda^2} = \frac{m_f}{v} + \frac{2 y_f' v^2}{2 \sqrt{2} \Lambda^2} \ ,
\label{eqn:Yukawa}
\end{equation}
are diagonal.  Simultaneous diagonalization of $m_f$ and $y_f'$ is not
guaranteed unless they are aligned, and hence without additional
assumptions, the Yukawa terms in dimension-6 Higgs effective theory
are expected to introduce flavor-changing Higgs couplings.  Moreover,
phases in $y_f'$ are not guaranteed to vanish, so we also expect $CP$
violation in Higgs couplings (the overall phase in each Yukawa matrix
is not observable).  Bounds on both flavor-changing Higgs couplings
and $CP$-violating couplings can be obtained from studying meson
mixing~\cite{Dery:2014kxa, Benitez-Guzman:2015ana} and electron and
neutron dipole moment constraints~\cite{Brod:2013cka}.

Nevertheless, a large, enhanced diagonal coupling for fermions is
readily achieved from~\eqnref{Yukawa}.  Note that for $y_u'$, $y_d'
\sim \text{ diag}(\mathcal{O}(1))$ and $v/\Lambda \sim
\mathcal{O}(1~\text{TeV})$, we obtain Yukawa enhancements $\kappa$ of
$\mathcal{O}(10^3-10^4)$ for first generation quarks,
$\mathcal{O}(10^2)$ for second generation quarks, and
$\mathcal{O}(10^2-10^{0})$ for third generation quarks, precisely
reflecting the universality of the dimension-6 Higgs $H^\dagger H /
\Lambda^2$ operator compared to the hierarchical structure of the SM
Yukawa matrix.

\section{$W^+ h$ vs. $W^- h$ charge asymmetry}
\label{sec:charge}

\begin{figure}[tb!]
\begin{center}
\includegraphics[scale=0.35, angle=0]{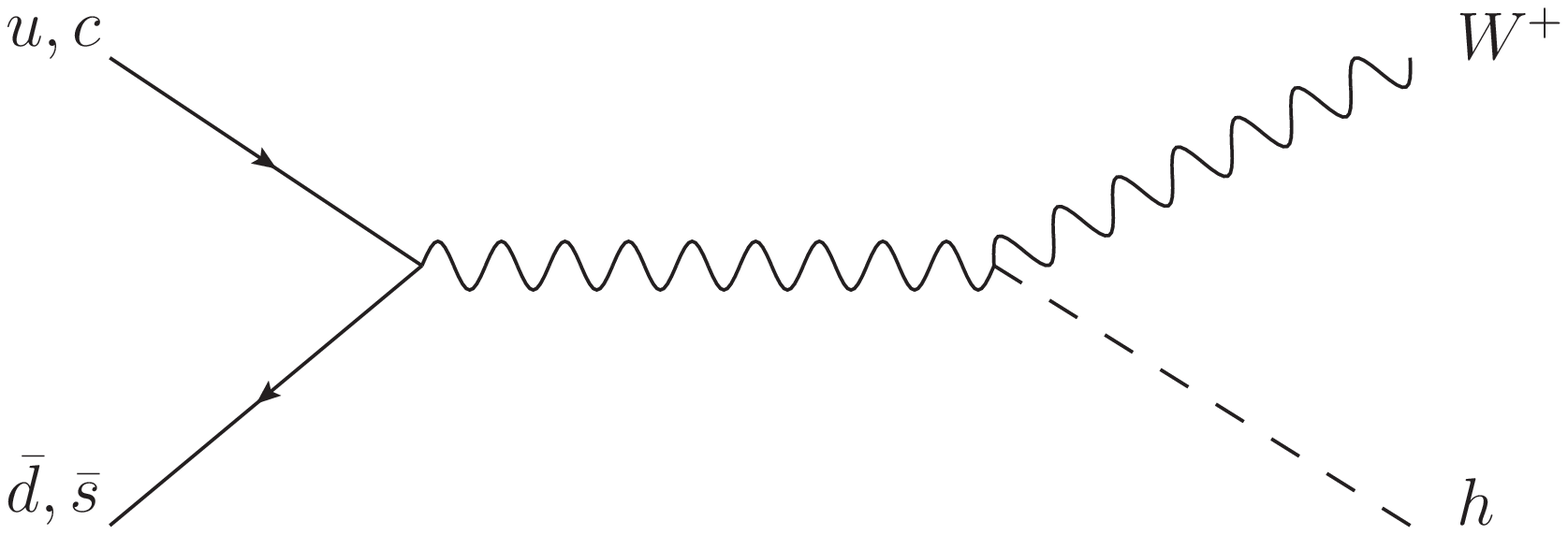}
\includegraphics[scale=0.35, angle=0]{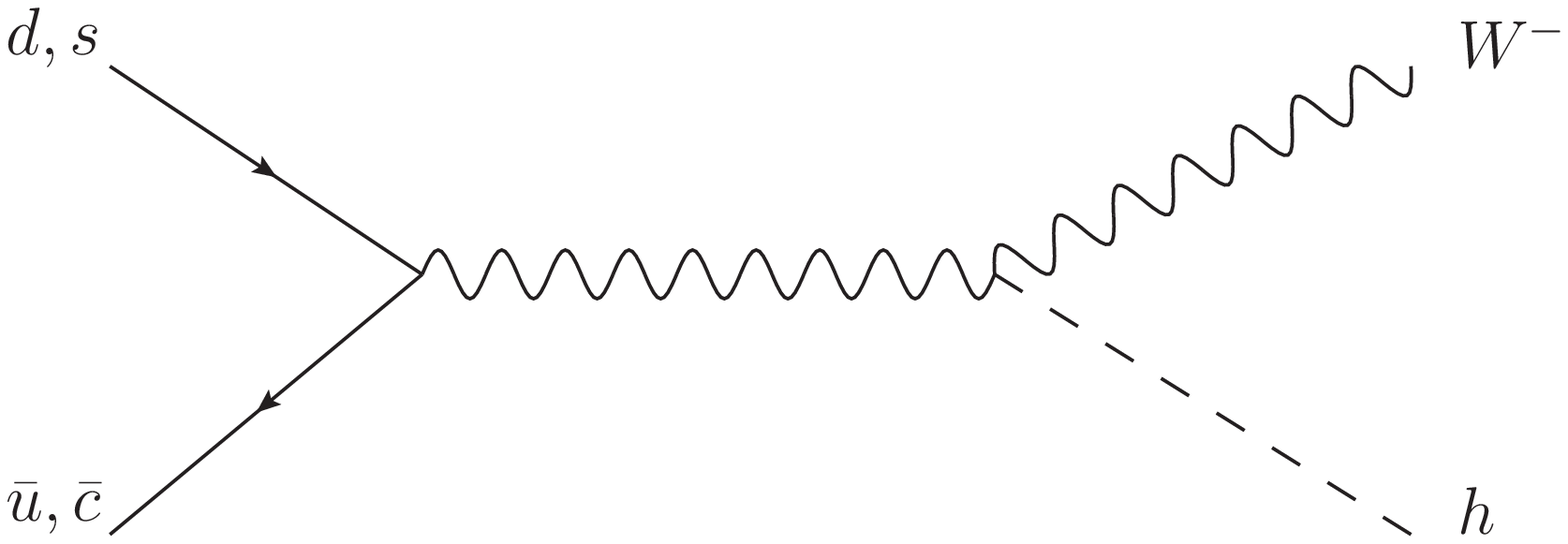} \\
\vspace{0.4cm}
\includegraphics[scale=0.40, angle=0]{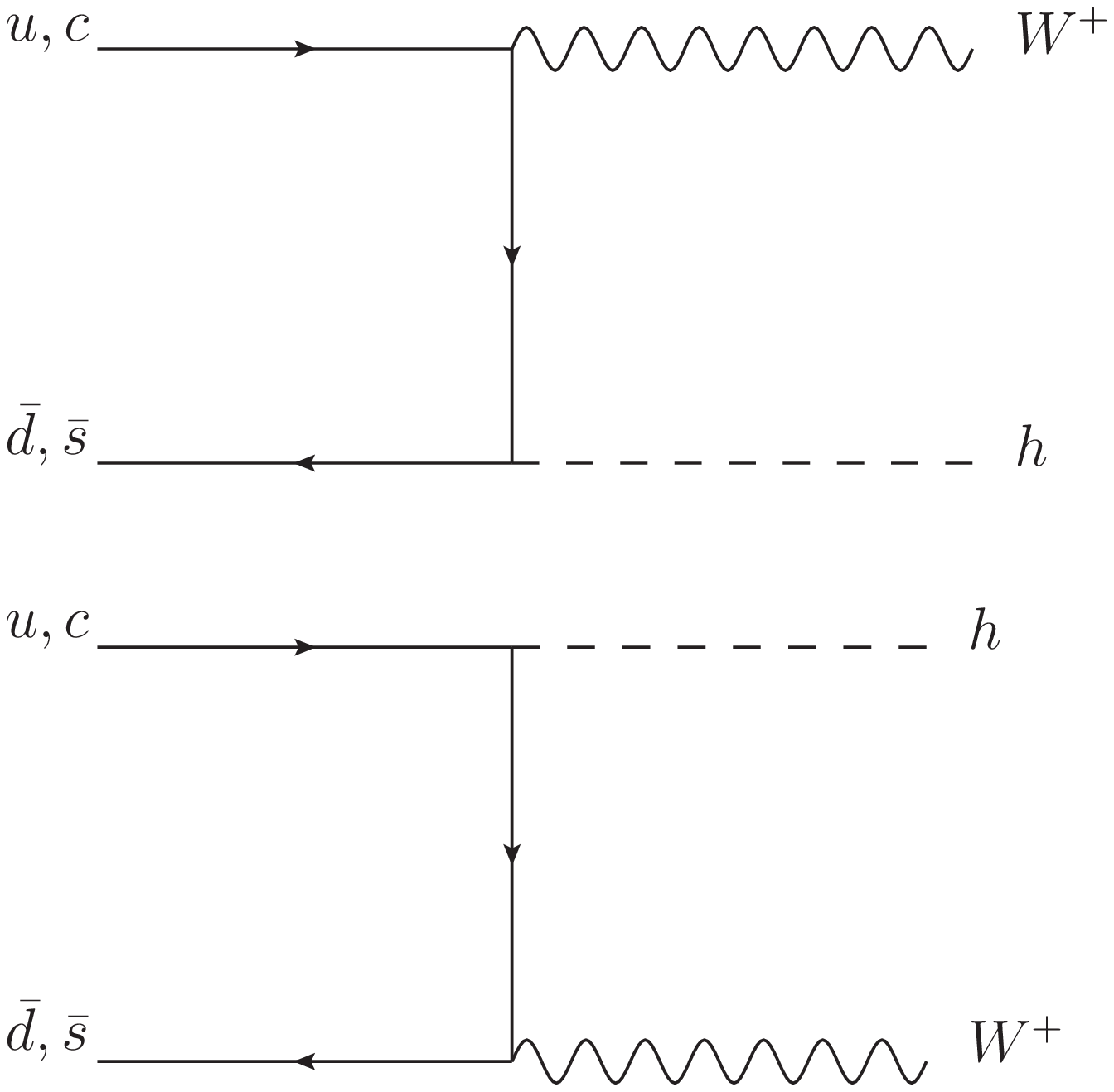}
\includegraphics[scale=0.40, angle=0]{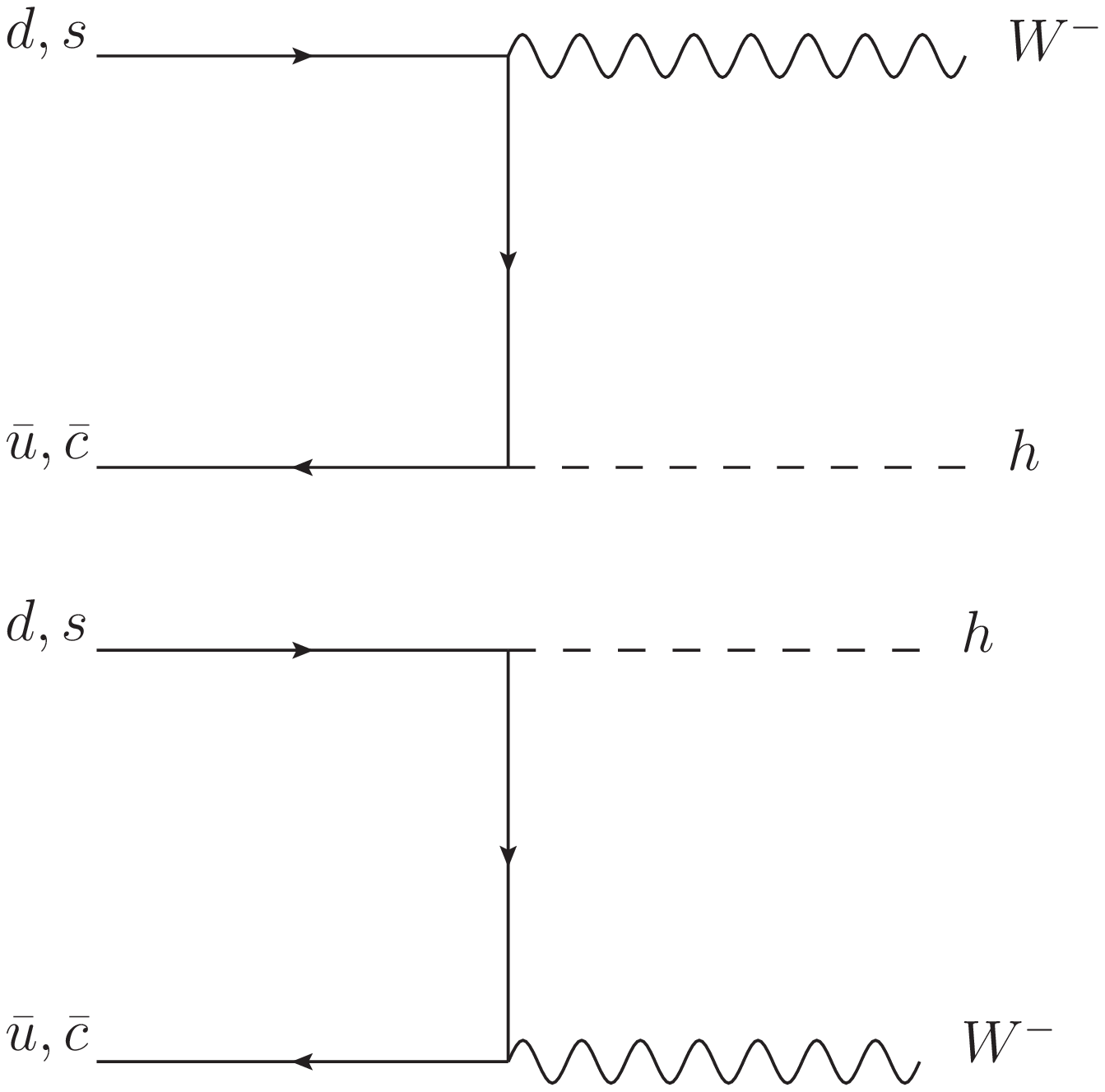}
\caption{Leading order $W^+ h$ (left column) and $W^- h$ (right
  column) production diagrams, showing the Higgsstrahlung process (top
  row) and Yukawa-mediated contributions (bottom two rows).}
\label{fig:diagrams}
\end{center}
\end{figure}

In the Standard Model, inclusive $W^\pm h$ production exhibits a
charge asymmetry of 21.8\% at the $\sqrt{s} = 14$~TeV
LHC~\cite{YellowReport, Heinemeyer:2013tqa}.  This charge asymmetry
directly results from the inequality of the LHC $pp$ parton
distribution functions (PDFs) under charge conjugation.  The tree
level diagrams for $W^\pm h$ production are shown
in~\figref{diagrams}, and in the SM, the Higgsstrahlung diagrams are
completely dominant compared to the Yukawa-mediated diagrams.  As a
result, the mismatch between $u \bar{d}$ vs.~$\bar{u} d$ PDFs at the
LHC drives the bulk of the charge asymmetry, which is ameliorated by
the more symmetric $c \bar{s}$ vs.~$\bar{c} s$ PDFs.  The
Cabibbo-suppressed contributions from $u \bar{s}$ vs.~$\bar{u} s$ and
$c \bar{d}$ vs.~$\bar{c} d$ PDFs also enhance and dilute,
respectively, the charge asymmetry.

Enhanced light quark Yukawa couplings cause the inclusive $W^\pm h$
charge asymmetry to deviate significantly from the SM expectation.
For very large Yukawa enhancements, we can neglect the Higgsstrahlung
diagrams in~\figref{diagrams} and focus on the Yukawa-mediated
diagrams.  If the charm Yukawa dominates the other couplings, then the
$c \bar{s}$ vs.~$\bar{c} s$ PDFs symmetrize $W^\pm h$ production, and
the overall charge asymmetry even turns negative from the residual $c
\bar{d}$ vs.~$\bar{c} d$ PDFs.  Similarly, an enhanced strange Yukawa
drives the balanced $c \bar{s}$ vs.~$\bar{c} s$ PDFs to dominate
$W^\pm h$ production, while the Cabibbo-suppressed $u \bar{s}$
vs.~$\bar{u} s$ initial states still retains a positive asymmetry.
Finally, large down and up quark Yukawas actually enhance the positive
charge asymmetry beyond the SM expectation, since the ameliorating
effects from second generation quarks in the proton PDFs are weakened.

We adopt the usual $\kappa$ notation to describe rescalings of the
Higgs Yukawa couplings to the first and second generation quarks,
$y_{f, \text{ eff}} = \kappa_f y_{f, \text{ SM}}$ for $f$ = $d$, $u$,
$s$, or $c$.  Throughout this work, we will only consider one Yukawa
deviation at a time and will comment briefly in the conclusions about
simultaneous deviations in multiple Yukawa couplings.  For
convenience, we also use the $\bar{\kappa}_f$ normalization, which
rescales $\kappa_f$ into units of $y_b^{\text{SM}}$ evaluated at $\mu =
125$~GeV:
\begin{equation}
\bar{\kappa}_f \equiv \dfrac{m_f (\mu = 125~\text{GeV})}{m_b (\mu =
  125~\text{GeV})} \kappa_f \ .
\label{eqn:kappabar}
\end{equation}

\begin{figure}[tb!]
\begin{center}
\includegraphics[scale=0.60, angle=0]{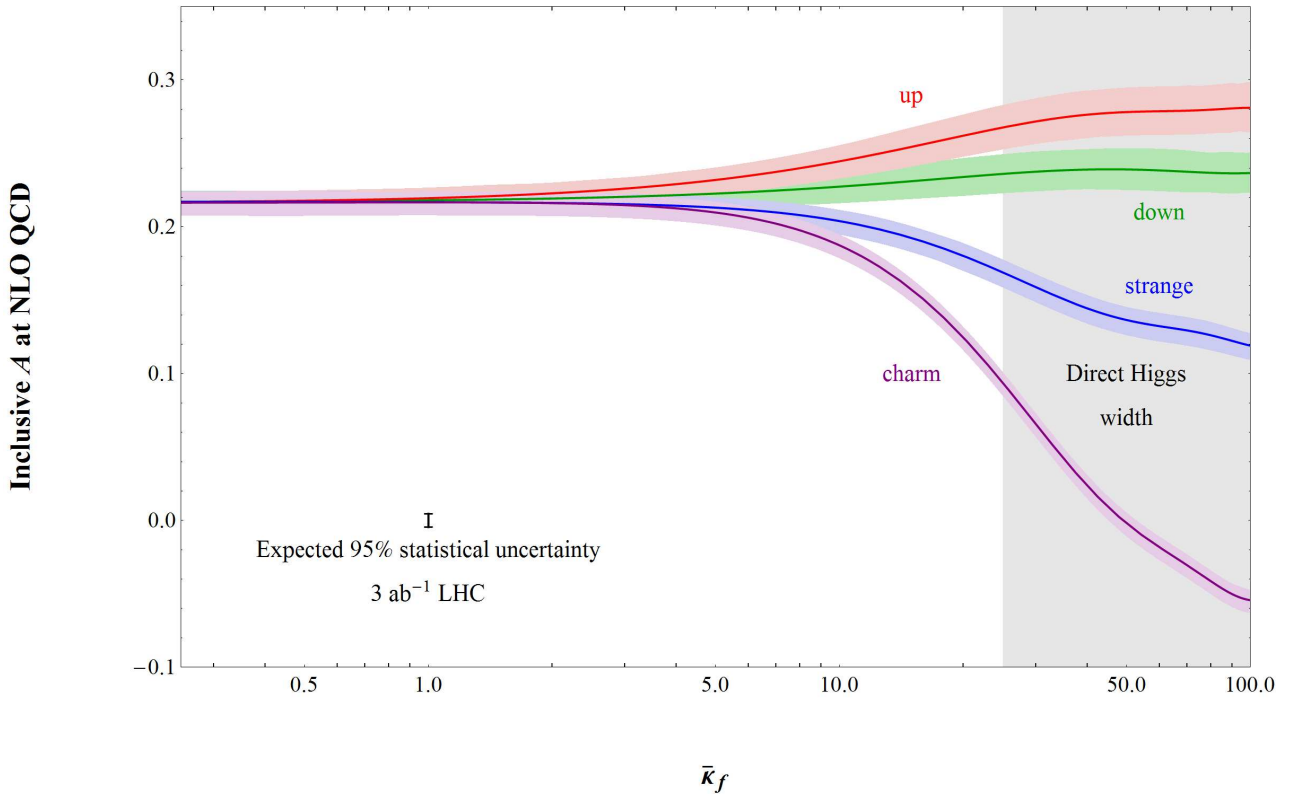}
\caption{Inclusive charge asymmetry $A = (\sigma(W^+ h) - \sigma(W^-
  h)) / (\sigma(W^+ h) + \sigma(W^- h))$ at NLO QCD for the $\sqrt{s}
  = 14$~TeV LHC as a function of individual Yukawa rescaling factors
  $\bar{\kappa}_f$ for $f = u$ (red), $d$ (green), $s$ (blue), and $c$
  (purple).  Shaded bands correspond to scale uncertainties at $1\sigma$
  from individual $\sigma(W^+ h)$ and $\sigma(W^- h)$ production,
  which are conservatively taken to be fully uncorrelated.  The gray
  region shows the bound from the direct Higgs width measurement,
  $\Gamma_H < 1.7$~GeV~\cite{Khachatryan:2014jba}, which excludes
  $\bar{\kappa}_f > 25$ for each light quark flavor and is discussed
  in~\secref{signalstrength}.  The expected statistical error from
  this measurement using 3 ab$^{-1}$ of LHC data is also shown.}
\label{fig:inclusivecharge}
\end{center}
\end{figure}

In~\figref{inclusivecharge}, we show the inclusive charge asymmetry
\begin{equation}
A = \dfrac{\sigma(W^+ h) - \sigma (W^- h)}{\sigma (W^+
h) + \sigma (W^- h)} \ ,
\end{equation}
for the 14~TeV LHC as a function of $\bar{\kappa}_f$ for individually
enhanced Yukawa couplings, $f = d$, $u$, $s$, and $c$.  These results
were generated using MadGraph v2.4.3~\cite{Alwall:2014hca} where the
Yukawa couplings were implemented via a
FeynRules~\cite{Alloul:2013bka} model implementing automatic
next-to-leading order (NLO) quantum chromodynamics (QCD) corrections
at 1-loop from NLOCT v1.0~\cite{Degrande:2014vpa} interfaced with
the NNPDF2.3 NLO~\cite{Ball:2010de} PDF set.  Yukawa couplings
were renormalized using the boundary values from the Particle Data
Group~\cite{Agashe:2014kda} and run to the Higgs mass with
RunDec~\cite{Chetyrkin:2000yt}.  The boundary values are $m_d =
4.8$~MeV, $m_u = 2.3$~MeV, $m_s = 0.95$~GeV at $\mu = 2$~GeV, and $m_c
= 1.275$~GeV at $\mu = m_c$.  We used a two-step procedure in the
renormalization group running to account for the change in the
$\alpha_s$ behavior at $b$-mass scale, $m_b = 4.18$~GeV at $\mu =
m_b$.  The extracted SM quark masses at $\mu = 125$~GeV are $m_d =
2.73$~MeV, $m_u = 1.31$~MeV, $m_s = 54$~MeV, $m_c = 634$~MeV, and $m_b
= 2.79$~GeV, which are used in~\eqnref{kappabar} to rescale $\kappa_f$
to $\bar{\kappa}_f$.  The Higgs coupling to $W$ bosons was fixed to
the SM value for this scan.

While QCD theory uncertainties are formally expected to cancel out in
a charge asymmetry, since QCD interactions respect charge
conservation, the factorization of the $W^\pm h$ partonic hard process
from the parent protons spoils this expectation and hence scale and
PDF uncertainties will not generally cancel.  We show the $1\sigma$
scale uncertainty for the whole range of $\bar{\kappa}_f$
in~\figref{inclusivecharge} as a shaded band.  We also evaluated the
PDF uncertainty using a leading order calculation interfaced with the
leading order NNPDF2.3 and CTEQ6L~\cite{Nadolsky:2008zw} PDF sets.
The two PDF sets leads to a $\approx 1\%$ disagreement in the
asymptotic values of the charge asymmetry for very large individual
$\kappa_f$.  

We remark that the statistical precision on the exclusive charge
asymmetry, which we propose to measure in~\secref{collider}, is
expected to be at the subpercent level, which we expect will improve
the overall status of PDF determinations at the
LHC~\cite{Rojo:2015acz}, regardless of the sensitivity to light quark
Yukawa couplings.  Moreover, $W^\pm h$ measurements complement $W^\pm
Z$ and $W^\pm + $ jets measurements, and improved measurements of the
charge asymmetry in these separate channels will confirm or refute
whether $W^\pm h$ production is dominated by the light quarks as
expected in the SM.

Measuring the asymmetry at the collider requires tagging the leptonic
decay of the $W$ boson and using a Higgs decay final state that
simultaneously tempers the background and retains sufficient
statistics to enable subpercent level accuracy.  In this vein, very
clean Higgs decays, such as $h \to Z Z^* \to 4\ell$ or $h \to \gamma
\gamma$ are inadequate for this purpose because the expected SM rates
for $\sigma(W^\pm h) \times $ Br$(W^\pm \to \ell^\pm \nu) \times $
Br$(h \to 4\ell)$ or Br$(h \to \gamma \gamma)$ are not statistically
large.  On the other hand, the largest SM Higgs decay channel, $h \to
b \bar{b}$, must contend with both the charge-symmetric semi-leptonic
$t \bar{t}$ background and the charge-asymmetric $W^\pm + $ jets
background: therefore, extracting the $W^\pm h$ charge asymmetry from
this Higgs final state will be challenging.  An interesting decay is
$h \to \tau^+ \tau^-$, where improvements in hadronic and leptonic
$\tau$ decays have led to important evidence for the Higgs decays to
taus~\cite{Khachatryan:2016vau}.  The efficacy of these Higgs
resonance reconstruction methods in the presence an additional lepton
and neutrino, however, has not been demonstrated.

We instead explore a new Higgs process, $W^\pm h \to (\ell^\pm \nu)
(\ell^\pm \nu j j)$, taking advantage of the semi-leptonic decay of
the Higgs via $WW^*$.  This process has a number of features
that make it attractive for measuring the $W^\pm h$ charge asymmetry.
First, this same-sign lepton final state inherits the same charge
asymmetry as the inclusive $W^\pm h$ process.  Second, the leading
non-Higgs background processes for same-sign leptons are all
electroweak processes, in contrast to the $h \to b\bar{b}$ decay
discussed before.  Finally, although the Higgs resonance is not
immediately reconstructible in this decay channel, we have a number of
kinematic handles to isolate the Higgs contribution to this final
state, which make it eminently suitable to extract the charge
asymmetry.

\section{Collider analysis: Same-sign leptons from associated $W^\pm h$ 
production}
\label{sec:collider}

Having motivated the possibility and importance of direct tests for
light quark Yukawa couplings via their effects in the charge asymmetry
of $W^\pm h$ production, we now present a search for $W^\pm h \to
\ell^\pm \ell^\pm \slashed{E}_T$ + 1 or 2 jets, with $\ell = e$ or
$\mu$, which can be a benchmark process for measuring the charge
asymmetry.  We emphasize that the charge asymmetry measured in an
exclusive Higgs decya mode is at best considered a consistency test of
the Standard Model, since large Yukawa deviations in light quark
couplings will dilute the SM Higgs branching fractions, which we
address in~\secref{signalstrength}.  Nevertheless, the charge
asymmetry of $W^\pm h$ production is a prediction of the Standard
Model that can be affected by deviations in light quark Yukawa
couplings.

The primary backgrounds for the $\ell^\pm \ell^\pm \slashed{E}_T$ + 1
or 2 jets signature are $W^\pm W^\pm jj$, $W^\pm Z$, with $Z \to
\ell^+ \ell^-$ and a lost lepton, and $W^+ W^-$ with charge
mis-identification.  Note that all of these diboson backgrounds are
electroweak processes, giving the benefit that $W^\pm h$ signal rates
are roughly comparable to the background rates.  On the other hand,
these backgrounds also have their own charge asymmetries, but these
can be probed via complementary hadronic channels, inverting selection
cuts, or data-driven techniques.

Other backgrounds we do not consider are fully leptonic $t\bar{t}$,
which we discard because it requires charge mis-identification and
would be killed by $b$-jet vetoes.  The single vector boson
backgrounds, $W + $ jets and $Z + $ jets, are neglected because they
need a jet faking a lepton or in the case of the $Z$ with charge
mis-identification, would still reconstruct the $Z$ peak.  We do not
consider hard brehmstrahlung with subsequent photon conversion, and we
ignore jet faking lepton rates, which eliminates QCD backgrounds.

Signal and background samples are generated for $\sqrt{s} = 14$~TeV
LHC using MadGraph 5 v2.2.1~\cite{Alwall:2014hca} at leading order in
QCD.  Signal bosons are decayed on-shell via $W^\pm \to \ell^\pm \nu$
and $h \to \ell^\pm \nu j j$, where the lepton charges are chosen to
be the same, and $\ell = e$ or $\mu$.  Backgrounds must pass the
preselection requirements of jet $p_T > 30$~GeV, lepton $p_T >
10$~GeV, and $\Delta R_{jj} > 0.2$.  In the background samples, $\tau$
leptons are included in the boson decays, since softer leptonic decays
from $\tau$s can contaminate the signal region.  We perform MLM
matching~\cite{Mangano:2001xp, Mangano:2002ea} for the $W^\pm Z$ and
$W^+ W^-$ backgrounds up to 1 jet, with the matching scale set to
30~GeV.  Events are passed to Pythia v6.4~\cite{Sjostrand:2006za} for
showering and hadronization and then simulated using a mock detector
simulation based on ATLAS and CMS performance measurements using
electrons~\cite{Aad:2011mk}, muons~\cite{ATLAS:2011hga},
jets~\cite{ATLAS:2011qda}, and
$\slashed{E}_T$~\cite{Chatrchyan:2011tn}.  We adopt an electron charge
mis-identification rate of 0.16\% for $0 < |\eta_e| < 1.479$ and 0.3\%
for $1.479 < |\eta_e| < 3$ and neglect muon charge
mis-identification~\cite{CMS-DP-2015-035}.

We calculate and apply flat NLO QCD $K$-factors using MCFM
v7.0~\cite{Campbell:1999ah, Campbell:2011bn, Campbell:2015qma} and
find $K = 1.71$ for $W^+ Z$, $K = 1.74$ for $W^- Z$, and $K = 1.55$
for $W^+ W^-$.  The NLO QCD corrections to the $W^\pm W^\pm jj$
background have been calculated in Refs.~\cite{Jager:2009xx,
  Melia:2010bm, Melia:2011gk}, from which we adopt a flat $K = 1.5$
factor.  After preselection, $K$-factors, and specified leptonic
branching fractions, our background rates are 113~fb for $W^\pm W^\pm
jj$, 630~fb for $W^+ Z$, 440~fb for $W^- Z$, and 8.80~pb for $W^+ W^-$.

To enhance the $W^\pm h$ contribution to the final state, we select
exactly two same-sign leptons with $p_T > 10$~GeV, $|\eta| < 2.5$.  We
then select either one or two jets with $p_T > 20$~GeV, $|\eta| <
2.5$, where jets are clustered using the anti-$k_T$
algorithm~\cite{Cacciari:2008gp} with $R = 0.4$ from FastJet
v3.1~\cite{Cacciari:2011ma}.  We allow events with only one jet
because the second jet from the Higgs decay is too soft or merges with
the first jet a significant fraction of the time.  Two-jet events are
required to be consistent with a hadronic $W$ candidate, $60$~GeV $<
m_{jj} < 100$~GeV.  Since the subleading lepton typically arises from
the Higgs semileptonic decay, we require $m_{T, \text{ subleading }
  \ell, jj} < 200$~GeV for two jet events.  These cuts are summarized
in~\tableref{cutflow}.

\begin{table}[tbh]
\renewcommand{\arraystretch}{1.2}
\begin{tabular}{|c|c|c|c|c|c|}
\hline
& {\bf SM $W^\pm h$} & {\bf $W^\pm W^\pm j j$} 
& {\bf $W^+ Z$} & {\bf $W^- Z$} & {\bf $W^+ W^-$} \\
Cross section, cut, survival efficiency 
& 6.5 fb + 4.2 fb & 113 fb & 630 fb & 440 fb & 8.80 pb \\
\hline
Exactly two leptons, $p_T > 10$ GeV & 
53.4\% & 32.6\% & 32.2\% & 31.9\% & 46.3\% \\
Same-charge leptons &
53.1\% & 31.7\% & 6.6\% & 6.6\% & 0.087\% \\
Either one or two jets, $p_T > 25$ GeV &
34.2\% & 22.5\% & 3.3\% & 3.4\% & 0.044\% \\
60 GeV $< m_{jj} < 100$ GeV & 
28.1\% & 11.7\% & 2.6\% & 2.6\% & 0.029\% \\
$m_{T,\text{ subleading } \ell jj} < 200$ GeV &
25.1\% & 4.9\% & 2.1\% & 2.2\% & 0.022\% \\
Number of events &
496 + 312 & 1070 + 604 & 3960 + 11 & 10 + 2860 & 270 + 303 \\
\hline
Statistical significance, 300 fb$^{-1}$, $S / \sqrt{S+B}$ &
\multicolumn{5}{c|}{6.5$\sigma$, 4.9$\sigma \Rightarrow 8.1\sigma$} \\
\hline
\end{tabular}
\caption{Cut flow for same-sign leptons from $W^\pm h$ production,
  where we denote the $++$ and $--$ contributions to the total number
  of events separately.}
\label{table:cutflow}
\end{table}

Normalizing the signal to the SM expectation~\cite{YellowReport,
  Heinemeyer:2013tqa}, $\sigma(W^+ h) \times $ Br$(W^+ \to \ell^+ \nu)
\times$ Br$(h \to \ell^+ \nu jj) = 6.5$~fb, $\sigma(W^- h) \times $
Br$(W^- \to \ell^- \bar{\nu}) \times$ Br$(h \to \ell^- \nu jj) =
4.2$~fb, where $\ell = e$ or $\mu$ only, we have a combined
statistical significance of $S / \sqrt{S + B} = 8.12\sigma$ from 300
fb$^{-1}$ of $14$~TeV LHC luminosity, and the individual $++$ and $--$
sign combinations are expected to reach $6.5\sigma$ and $4.9\sigma$,
respectively.  Hence, this mode should provide practical discovery
sensitivity to $W^\pm h$ production compared to the null hypothesis.
Although this mode does not admit a resonant reconstruction of the
Higgs candidate, the presence of the two same-charge leptons with
manageable background rates makes it a uniquely robust analysis for
studying the $W^\pm h$ charge asymmetry.

After our cuts, the $W^\pm h$ signal asymmetry is 22.8\%, while the
total charge asymmetry from background contamination is 16.8\%.  A
more careful study of systematic effects, subleading backgrounds, and
further reduction of the diboson backgrounds in this channel is
certainly warranted but beyond the scope of this work.  Optimized cuts
on the hadronic $W^{(*)}$ candidate from the Higgs signal would help
minimize the dominant charge-asymmetric $W^\pm Z$ background and
improve the signal to background discrimination.  Moreover, binning
the charge asymmetry according to the leading lepton pseudorapidity
can help test enhanced Yukawa couplings via different admixtures of
underlying PDF contributions.  We note, however, that this is a
relatively mild effect because the leading lepton does not always
originate from the associated parent $W$ and, in the case that one
Yukawa coupling is enhanced at a time, the signal $W^\pm h$ process is
still produced from combinations of valence and sea quark PDFs because
of the non-negligible Cabibbo angle.

We expect future studies from additional reconstructable decay modes
of the Higgs, such as $h \to b \bar{b}$, $h \to \ell^+ \ell^- \nu \nu$
(via $Z Z^*$ or $W W^*$), $h \to \tau^+ \tau^-$, and $h \to \gamma
\gamma$ will also contribute to the overall sensitivity of measuring
the $W^\pm h$ charge asymmetry.  Each of these modes requires,
however, a dedicated discussion of the charge asymmetries in their
dominant backgrounds, which is the scope of future work.  We expect
that decay channels giving comparable discovery significance of the
$W^\pm h$ associated production mode will add further improvements to
an overall global fit of PDFs, if $W^\pm h$ production is assumed to
be SM-like, and simultaneously, be instrumental in testing for
nonstandard light quark Yukawa couplings.

Extrapolating to $3$ ab$^{-1}$, we find that the charge asymmetry of
the $W^\pm h$ process can be tested with a statistical precision of
$\approx 0.4\%$, which would be sensitive to higher order theory
uncertainties, including PDF errors, and experimental systematic
uncertainties, which we have neglected in this treatment.  We note
that the statistical precision will be comparable to the QCD scale
uncertainty in the theory calculation and the expected
$\mathcal{O}(few)\%$ PDF uncertainty already with $300$ fb$^{-1}$ of
LHC luminosity using the current cuts.  Rigorous optimization of this
analysis focusing on improving the signal to background ratio,
however, will avoid this sensitivity saturation to larger
luminosities.  Moreover, improved measurements of charge asymmetries
in $W + $ jets and $W^\pm Z$ processes~\cite{Rojo:2015acz} will
further reduce the light quark PDF uncertainties, while measurements
of $W^\pm c$ rates with charm-tagging will significantly improve the
determination of $s$, $\bar{s}$ PDFs.  Overall, the charge asymmetry
in the $W^\pm h$ channel complements the charge asymmetry measurements
in other $W^\pm$ production modes, and in the event of a discrepancy,
provides a direct, diagonstic tool to test for enhanced Higgs
couplings to light quarks.

We remark that for non-standard Yukawa couplings, the kinematic
distributions for $W^\pm h$ production are expected to change,
resulting in small differences in the quoted efficiencies.  For
example, with $\kappa_d = 1000$ ($\kappa_u = 1000$) the final $W^\pm
h$ signal efficiency decreases to 24.8\% (24.5\%) compared to the SM
benchmark efficiency of 25.1\%.

\section{Phenomenology of Quark Yukawa couplings and current constraints}
\label{sec:signalstrength}
The set of Higgs measurements from the LHC and the Tevatron provide a
broad but patchwork picture of Higgs couplings constraints.  We
emphasize that a direct measurement of Higgs couplings at the LHC is
not currently feasible since the total width of the Higgs is unknown,
and thus interpretation of Higgs measurements requires model assumptions
about the underlying Lagrangian dictating the Higgs couplings and
possible new light degrees of freedom.  For example, the
$\kappa$-framework for studying Higgs coupling deviations is invalid
when new exotic modes for Higgs production are
accessible~\cite{Yu:2014mda}, which cause changes in signal efficiency
that are not captured by simple coupling rescalings.

\subsection{Total width constraints}
\label{sec:total}
The only direct test for enhanced light quark Yukawa couplings from
the LHC is the constraint from the direct measurement of the total
Higgs width.  From the 7+8~TeV combined analyses using the $\gamma
\gamma$ and $4\ell$ channels, ATLAS reported a Higgs total width
$\Gamma_H$ constraint of 2.6~GeV at 95\% CL~\cite{Aad:2014aba} and CMS
reported a tighter bound of 1.7~GeV~\cite{Khachatryan:2014jba}.  With
the latest 13~TeV data, CMS observed a bound of 3.9~GeV (expected
2.7~GeV) in the $4\ell$ channel~\cite{CMS:2016ilx} compared to a bound
of 3.4~GeV (expected 2.8~GeV) with the Run I
dataset~\cite{Chatrchyan:2013mxa}, indicating that lineshape
measurements of the Higgs have already saturated the resolution
expected from the LHC.  We remark that the next-generation $e^+ e^-$
Higgs factory machines~\cite{Baer:2013cma, Gomez-Ceballos:2013zzn,
  CEPC-SPPCStudyGroup:2015csa} will inaugurate the true precision era
of Higgs measurements by virtue of being able to tag Higgs-candidate
events via the recoil mass method, which can determine the SM Higgs
width with 2--5\% precision~\cite{Dawson:2013bba}.  Since light
quarks are kinematically accessible decay modes of the 125~GeV Higgs,
however, the on-shell decay of the Higgs to light quarks via enhanced
Yukawa couplings is untamed for large $\bar{\kappa}$.

We can thus use the $\Gamma_H < 1.7$~GeV constraint from
CMS~\cite{Khachatryan:2014jba} to bound the individual light quark
Yukawa couplings:
\begin{equation}
\kappa_d < 27500, \quad \kappa_u < 57400, \quad \kappa_s < 1300, \quad
\kappa_c < 120 \ ,
\end{equation}
using the renormalized quark masses calculated from
RunDec~\cite{Chetyrkin:2000yt}.  These translate to
\begin{equation}
\bar{\kappa}_f \lesssim 25 \ ,
\end{equation}
for each of the first or second generation light quarks, $f = d$, $u$,
$s$, or $c$.  These bounds are indicated in the gray region
of~\figref{inclusivecharge}.

If we recast the latest indirect measurements of the Higgs width
$\Gamma_H < 41$~MeV~\cite{CMS:2016ilx}, obtained from ratios of
Higgs-mediated events in $gg \to ZZ \to 4\ell$ production in off-shell
vs.~on-shell Higgs regions~\cite{Kauer:2012hd, Caola:2013yja,
  Campbell:2013una}, we find $\bar{\kappa}_f \lesssim 4$.  This bound
depends, however, on model assumptions about the behavior of Higgs
couplings in the off- and on-shell regions, controlled theory
uncertainties in the NLO QCD corrections to the interference between
the $gg \to ZZ$ box diagram and the Higgs amplitude, and fixing all
other Higgs partial widths to their SM values.  Referring
to~\figref{inclusivecharge}, this current bound still permits a
percent-level deviation in the inclusive charge asymmetry, which we
expect is measureable with the full dataset of the LHC.  In our view,
the indirect width measurement of the Higgs and the charge asymmetry
measurement are equally valid as consistency tests of the Standard
Model Higgs, and we strongly advocate for the charge asymmetry test in
future LHC Higgs analyses.

\subsection{Inclusive charge asymmetry}
\label{sec:inclusiveasym}
At the fully inclusive level, the Higgs Yukawa couplings can be tested
via the proposed charge asymmetry measurement.  While more stringent
constraints on the light quark Yukawa couplings can be obtained from
global fits combining all Higgs data, these global fits suffer from
the requirement of a theoretical model dependence, most commonly the
$\kappa$ framework.  

We point out, however, that absent deviations in light quark Yukawa
couplings the fully inclusive charge asymmetry also provides a
model-independent measurement of the Higgs coupling to $W$ bosons.
Fully inclusive Higgs production processes are not normally considered
at hadronic colliders because of the inability to ascertain the Higgs
contribution independent of the Higgs decay mode.  This is analogous
to the recoil mass method advocated for $e^+ e^-$ Higgs factories,
which allows a fully inclusive rate measurement sensitive to the $hZZ$
coupling.  At the moment, though, there is no practical proposal for
measuring such an inclusive variable in any Higgs process and all
Higgs data stems from analyses for specific Higgs decays, and so the
intriguing possibility of a fully inclusive Higgs measurement to
extract a Higgs production coupling remains remote.

\subsection{Exclusive Higgs measurements and current constraints}
\label{sec:exclusive}

In~\eqnref{Lagrangian}, we only introduced new physics operators that
modified the mass generation and Yukawa couplings of the SM quarks,
leaving the Higgs-vector couplings untouched.  As a result, enhanced
Yukawa couplings lead to increased rates for $\sigma (q q' \to W^\pm
h)$ and $\sigma (q \bar{q} \to h)$ production, but the effective
signal strengths $\mu_{Wh}$ and $\mu_{gg}$ of exclusive Higgs decays
to a particular $X$ final state are depleted according to
\begin{eqnarray}
\mu_{Wh} (h \to X) &=& \dfrac{ \left(\sigma_{Wh}^{\text{NP}} \right)}{
  \left(\sigma_{Wh}^{\text{SM}} \right)} \times \dfrac{ \Gamma (h \to
  X)^{\text{NP}} / \Gamma_H^{\text{NP}}}{\Gamma (h \to
  X)^{\text{SM}} / \Gamma_H^{\text{SM}} } \ , \\
\mu_{gg} (h \to X) &=& \dfrac{ \left(\sigma_{gg}^{\text{NP}} +
  \sigma_{qq}^{\text{NP}} \right)}{ \left(\sigma_{gg}^{\text{SM}}
  \right)} \times \dfrac{ \Gamma (h \to X)^{\text{NP}} /
  \Gamma_H^{\text{NP}}}{\Gamma (h \to X)^{\text{SM}} /
  \Gamma_H^{\text{SM}} } \ ,
\label{eqn:signalstrength}
\end{eqnarray}
where we have included $s$-channel $q\bar{q}$ Higgs production in the
overall gluon fusion rate.  We remark that the gluon fusion and
$q\bar{q}$ annihilation production modes can be possibly disentangled
at the LHC by studying Higgs candidate
kinematics~\cite{Bishara:2016jga, Soreq:2016rae, Bonner:2016sdg},
while the $q\bar{q}$ decay can also possibly be probed at $e^+ e^-$
Higgs factories~\cite{Gao:2016jcm}.

Solely turning on large Yukawa couplings for light quarks is hence
strongly constrained by combined coupling fits using current Higgs
data, since the increased production rates from the Yukawa-mediated
processes is not enough to counterbalance the rate loss in measured
Higgs modes such as $h \to 4 \ell$ and $h \to \gamma \gamma$.  For
example, if we require that $\mu_{gg} (h \to 4\ell)$ is within 40\% of
the SM signal strength, consistent with the latest 13~TeV Higgs
measurement results~\cite{CMS:2016ilx} and only allow one light quark
Yukawa coupling to deviate at a time, then we derive the following
constraints:
\begin{equation}
\kappa_d < 1270, \quad \kappa_u < 2860, \quad \kappa_s < 53, \quad
\kappa_c < 5 \ ,
\label{eqn:naive}
\end{equation}
which can be converted to
\begin{equation}
\bar{\kappa}_d < 1.24, \quad \bar{\kappa}_u < 1.34, \quad 
\bar{\kappa}_s < 1.03, \quad \bar{\kappa}_c < 1.14 \ ,
\end{equation}
where we have fixed $\sigma_{gg} = 48.58$ pb~\cite{Anastasiou:2014vaa,
  Anastasiou:2016cez} using $m_H = 125$~GeV for both the SM and NP
rates and only considered the additional contribution from $q\bar{q}$
annihilation.  These ad-hoc constraints are only presented to
demonstrate the naive sensitivity to light quark Yukawa couplings from
a 1-parameter test, where all other SM couplings are held fixed.  We
note that the intrinsic contribution from light quarks affecting gluon
fusion is suppressed by the loop function dependent on the quark
masses.  Moreover, new colored particles in the gluon fusion loop
(see, e.g., Ref.~\cite{Kumar:2012ww} and references therein) can add
to the $s$-channel $q\bar{q}$ Higgs production channel to compensate
for the drop in the $h \to 4 \ell$ branching fraction.  In principle,
an enhanced coupling of the Higgs bosons to electroweak vectors can
also relieve the bounds above, although concrete possibilities are
limited~\cite{Logan:2015xpa}.  A global analysis performed in
Ref.~\cite{Kagan:2014ila}, allowing all Higgs couplings to vary, has
derived the constraints $\bar{\kappa}_d < 1.4$, $\bar{\kappa}_u <
1.3$, $\bar{\kappa}_s < 1.4$, and $\bar{\kappa}_c < 1.4$.

We note that the Tevatron also provides constraints on enhanced light
quark Yukawa couplings given the nature of the machine as a
proton--anti-proton collider.  The primary search channel at the
Tevatron sensitive to $s$-channel Higgs production was the $WW^*$
decay mode~\cite{Aaltonen:2013ioz}, which constrained $\sigma (gg
\rightarrow H) \times$ Br$(H \to WW^*)$ at $m_H = 125$~GeV to be less
than 0.77~pb.  If $\sigma (gg \to H)$ and Br$(H \to WW^*)$ are held
fixed, then this constrains the extra production from $\sigma
(q\bar{q} \to H)$ at a level roughly a factor of 2-10 weaker than the
naive estimate in~\eqnref{naive}, with the strongest constraints for
$\kappa_d$ and $\kappa_u$; again, this is an inconsistent treatment of
the bounds unless new physics is introduced to keep Br$(H \to W^+
W^-)$ fixed.  In a similar manner, double Higgs production rates are
also increased, but their impact at the LHC is already excluded in a
model independent fashion from the total Higgs width measurement
discussed earlier.

Finally, probing enhanced quark Yukawa couplings using the exclusive
charge asymmetry measurement discussed in~\secref{collider} requires
also requires an increased $h \to WW^*$ partial width in order to
maintain the signal rate comparable to the SM expectation.
Nevertheless, the measurement of the charge asymmetry provides an
important consistency test of the SM Higgs boson.  Moreover, the
$0.4\%$ statistical precision afforded by the proposed $W^\pm h \to
\ell^\pm \ell^\pm jj + \slashed{E}_T$ measurement establishes a new
channel to constrain and evaluate parton distribution functions and
their uncertainties if light quark Yukawa deviations are absent.

\section{Conclusions}
\label{sec:conclusions}

In this work, we have explored the prospects for measuring light quark
Yukawa couplings at the LHC via the charge asymmetry of $W^\pm h$
production.  From the limited set of new physics operators considered,
the net effect of enhanced light quark Yukawa couplings was to rapidly
increase the total Higgs width, which can be tested in a
model-independent fashion at the LHC in the high resolution $\gamma
\gamma$ and $4\ell$ final states.  Enhanced light quark Yukawa
couplings consistent with the direct Higgs width constraint predict
inclusive charge asymmetries that deviate significantly from the SM
expectation.

We hence motivated the possible measurement of the $W^\pm h$ charge
asymmetry in the exclusive mode $W^\pm h \to \ell^\pm \ell^\pm
\slashed{E}_T$ + 1 or 2 jets, which is a clean same-sign dilepton
final state that inherits the same charge asymmetry as the original
Higgs production process.  After accounting for the main backgrounds
from electroweak diboson production, we estimate that the individual
$++$ and $--$ final states reach a statistical $\approx 5\sigma$ significance
each with 300 fb$^{-1}$ of 14~TeV LHC data.  Even though the Higgs
boson is not fully reconstructed in this decay, the clean same-sign
dilepton signature can be readily extrapolated to the expected $3$
ab$^{-1}$ high luminosity run, enabling a statistical precision on the
exclusive charge asymmetry of 0.4\%.  If the measured asymmetry
deviates from the SM expectation, then a likely interpretation would
be an enhanced SM light quark Yukawa counterbalanced by additional new
physics effects that preserve rough current consistency of the Higgs
data with SM expectation.  A future deviation can favor enhanced down
and up quark Yukawas if the observed charge asymmetry exceeds the SM
expectation, while strange and charm quark Yukawas would be
responsible if the charge asymmetry were smaller.  

The $W^\pm h$ charge asymmetry hence provides an interesting and new
consistency test for Higgs measurements.  We conclude by remarking
that although we focused on the prospects for testing light quark
Yukawa coupling deviations using the charge asymmetry, this
measurement also probes the Higgs coupling to $W^\pm$ bosons directly,
which adds a new ingredient in combined coupling fits for testing
custodial symmetry.

\section*{Acknowledgments}
\label{sec:acknowledgments}
The author is grateful to Wolfgang Altmannshofer, Fady Bishara,
Joachim Brod, Maikel de Vries, Stefan Kallweit, Joachim Kopp, Andreas
von Manteuffel, Gilad Perez, Yotam Soreq, and Nhan Tran, for useful
discussions.  This research is supported by the Cluster of Excellence
Precision Physics, Fundamental Interactions and Structure of Matter
(PRISMA-EXC 1098), by the ERC Advanced Grant EFT4LHC of the European
Research Council, and by the Mainz Institute for Theoretical Physics.
The author is grateful to the Universit\'{a} di Napoli Federico II and
INFN for its hospitality and its partial support during the completion
of this work.  The author would also like to acknowledge the
hospitality of the respective theory groups from the IBS CTPU in
Korea, the Technische Universit\"{a}t Dortmund, the Technion, and the
Tel Aviv University, where parts of this work were completed, as well
as thank the organizers and participants of the Higgs Effective Field
Theories 2016 (HEFT2016) in Chicago for their stimulating comments and
discussion.




\bibliographystyle{apsrev4-1}

\end{document}